# Analysis of Compression Techniques for DNA Sequence Data

SHAKEELA BIBI
JAVED IQBAL
ADNAN IFTEKHAR
MIR HASSAN
School of Computer Science
Wuhan University, China

**Abstract:**

*Biological data mainly comprises of Deoxyribonucleic acid (DNA) and protein sequences. These are the biomolecules which are present in all cells of human beings. Due to the self-replicating property of DNA, it is a key constitute of genetic material that exist in all breathing creatures. This biomolecule (DNA) comprehends the genetic material obligatory for the operational and expansion of all personified lives. To save DNA data of single person we require 10CD-ROMs. Moreover, this size is increasing constantly, and more and more sequences are adding in the public databases. This abundant increase in the sequence data arise challenges in the precise information extraction from this data. Since many data analyzing and visualization tools do not support processing of this huge amount of data. To reduce the size of DNA and protein sequence, many scientists introduced various types of sequence compression algorithms such as compress or gzip, Context Tree Weighting (CTW), Lampel ziv Welch (LZW), arithmetic coding, run length encoding and substitution method etc. These techniques have sufficiently contributed to minimize the volume of the biological datasets. On the other hand, the traditional compression techniques are also not much suitable for the compression of these types of sequential data. In this paper, we have explored diverse types of techniques for compression of large amount of DNA Sequence data. In this paper, the analysis of techniques reveals that efficient techniques not only reduce the size of the sequence but*

*also avoid from any information loss. The review of existing studies also shows that compression of DNA sequence is significant for understanding the critical characteristics of DNA data in addition to improving storage efficiency and data transmission. In addition, the compression of protein sequence is a challenge for the research community. The major parameters for evaluation of these compression algorithms includes compression ratio, running time complexity etc.*

**Key words:** Huffman Coding, Extended ASCII Illustration, DNA Encryption, Decompression

## I. INTRODUCTION

Due to various genome projects, noteworthy amount of DNA bases of different species are placed in many databases such as PCR and Quantitative PCR Primer Databases, Taxonomic Databases, Genome Databases etc. Human DNA has 3 billion base pairs. It is the major carrier of genetic information and there are 20,000-25000 genes within DNA. Long DNA base pair sequence data can be represented graphically for better understanding by using Huffman Coding. Compression of DNA Sequence data has been of interest for many decades. DNA is organized in chromosomes which are located in the center of cells. The nucleus of hominid cells comprises 46 chromosomes, each of which contains a particular lined molecule of deoxyribonucleic acid (DNA), which are confidentially multiplexes with proteins in the form of chromatin. DNA is the building block of life, which comprehends programmed genetic information for breathing creatures. A DNA is transliterated to convert a predecessor mRNA, which is then interweaved to become an mRNA, which is interpreted to become a protein. Because, except some cells all the cells in a humanoid physique comprehend an indistinguishable set of genes, for different cells of human body the appearance level of every gene must be different [1].

It is a major issue to store large amount of DNA sequence data which consist of long chain of DNA base pair sequence data. DNA and protein sequence data bases amount to hundreds of gigabyte (GB) storage space. This makes data compression in genomics a very important and challenging task. For storing the DNA and protein sequence, publicly available databases are EMBL, Gene bank, and DDBJ. The size of these databases is increasing exponentially [2].

Recently, many scientists have introduced different type of algorithms to solve the compression problem of genetic sequence data. In 1983 Hamori planned Graphical representation aimed at visualizing the DNA classification initially [3]. After this Nandy introduced a graphical method by assigning four bases of DNA A, C, T and G in the different four directions, like (x),(y),(-x),(-y) correspondingly in 1994. This algorithm visualized the DNA arrangement that is directed by certain cost of graphic material associated through crossing then covering of the curve by it while it bounces with abundant natural symbol for extended DNA categorization [4].

The results of earlier mentioned algorithms after compression have been compared with DNA sequence data of 10 distinct species. Graphical approaches have great importance for the breakdown and conception of long DNA structures. Different authors have presented few additional graphical methods of DNA classifications in 2Dimensional, 3Dimensional, and 4Dimentional planetary. Graphical symbol of long DNA sequence are too helpful for inspecting, categorization and comparing dissimilar genes construction. Graphical illustration of a DNA sequence delivers valuable visions into native and comprehensive features towards a sequence, which is not easy to observe from a DNA sequence. Multicomponent vectors derived after a DNA arrangement is hands-on in creation resemblances/differences of arrangements [5].

DNA sequence data has repetition of information. As we are familiar that recurrent DNA base pair arrangements are accountable for organic specialization proposed by Grechko in 2011 and palindromic structures relate to spots of DNA breakage throughout the gene transformation. Therefore, it's not easy for bio scientist to manage the repeated large amount of DNA sequence data so that compression techniques have been introduced to decrease the extent of redundant data and appropriately examine which data is important and should be stored. Discovery of recurring structures has been an elementary step for refining a material structure presentation and falling the quantity of the requiring space. Firmness of the twofold output of DNA datasets to extended ASCII illustration is shown in Figure 1

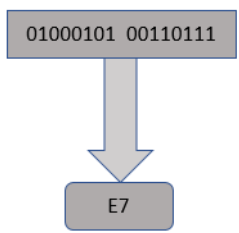

**Figure 1: ASCII representation [5]**

Huffman coding algorithm is a lossless compressing algorithm that compresses data without any cost of its statistics. Consequently, it is used for text firmness techniques and therefore for DNA arrangement data. In this paper, the authors explained a graphical demonstration of DNA main classification founded on Huffman coding algorithm, in this algorithm, there is no immorality of material of DNA sequence. DNA sequence analysis is supportive in diverse areas as forensics, medical research, pharmacy, agriculture etc. [6].

      However, it is proved that compression algorithms that are used for text compressions are not efficient for compression of DNA sequences. Therefore, there are two types of compression algorithms. One is condensing in horizontal mode

and second is condensing in vertical mode. In the first one compression is held on solo sequence based on its genetic materials as Biocompress [7] pursues replications in a sequence data. When this method is applied on standard benchmark datasets then it shows 1.85 BpB compression ratios. To compress the non-reparative part of sequence Biocompress-2[7] Markov model is used, and it shows compression ratio of 1.78 BpB.

This paper is summarized as follows. Section II provides literature review of the related work; Section III presents the main problem and objective of this research work. In Section IV, we proposed comparison of different tools and techniques that are described in detail by different authors in different papers. Results are presented and discussed in the section V. Conclusion is provided in the last section.

## II. LITERATURE REVIEW

Various DNA compression techniques have been presented recently, a literature review of few of those techniques is presented in the following section.

Bacem Saada et al, proposed a DNA sequences data compression by using Huffman Coding in 2016. In their paper, Authors introduced two levels compression process founded on the binary illustration of DNA base pair arrangements. Authors used lossless compression method for compression of nucleotides. This algorithm use statistical approach which works in 2 steps. In the first part, Authors implemented variation of Huffman coding to easily cover and renew the DNA base pair categorization into twofold (paired) representation. Authors have introduced the technique to compress the DNA sequence data which is based on binary representation of DNA sequence data. The key objective of the primary step of this technique is to allocate variable-length that encrypts the input chunks of the DNA arrangements. This phase of procedure

Shakeela Bibi, Javed Iqbal, Adnan Iftekhar, Mir Hassan- **Analysis of Compression Techniques for DNA Sequence Data**

implemented as follows AAAC, ACGT, GAGA, ACGT, TAAC, GATC, ACGT, TAAC. Table 1 shows the DNA Words and their occurrence frequency.

**Table 1: Vocabulary Table Content [5]**

| Index | Word | Frequency |
|---|---|---|
| 0 | ACGT | 3 |
| 1 | GAGA | 1 |
| 2 | GATC | 1 |
| 3 | TAAC | 2 |
| 4 | AAAC | 1 |

The distances of the allocated codes are according to the number of incidences of the chunks. The most common chunk is allocated the least code and the less common chunk is allotted the main code [5]. Table 2 illustrates the words and relevant code to be used in the example data.

**Table 2: Vocabulary Table Codes [5]**

| Index | Word | Frequency |
|---|---|---|
| 0 | ACGT | 00 |
| 1 | TAAC | 10 |
| 2 | GATC | 0001 |
| 3 | GAGA | 0010 |
| 4 | AAAC | 1101 |

The Results of arrangement will be as 00 00 0001 11 00 11 0001 1101

After this to efficiently decrease the scope of the output consequence by using Extended-ASCII programming they compressed the DNA sequence data concluded that which one block can signify 16 bytes. This technique shows the significant compression ratio and easy to understand and implement. The main objective of this method is that one extended-ASCII charm encrypts eight binary numbers. Moreover, Figure 2 shows the additional data structure employed for the output sequence data.

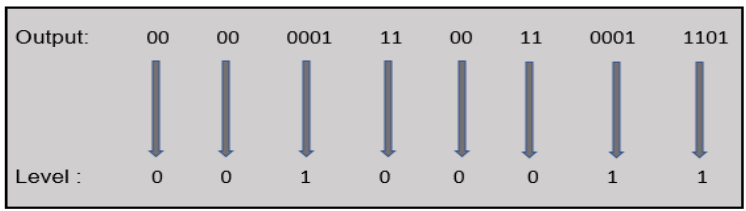

**Figure 2: Additional Data Structure [5]**

The output consequence will be condensed to 12.5% of its original representation. The efficiency of this technique is that it has 0.8 BpB compression ratio per base which is better and efficient than all other currently existing compression algorithms [8]. For this technique, authors used the HUMHBB, HUMHDABCD, MPOMTCG and VACCG which have size of more than 180000 nucleotides [5].

In 2016 George Volis et al, has introduced dual innovative techniques for space compaction of biological sequence data. In this work, authors discussed the problem of management of large amount of DNA sequence data. Authors introduced two techniques that are based on n-grams/2L method for compression of DNA sequence data. Authors proposed the solution of eliminating the size of memory for saving the genetic material. The first method is founded on 2L technique that indexed the DNA sequence data for better understanding and compression and it can change the upturned catalog of this standard procedure to an extra flattened presentation. Figure 3 shows the comparisons results achieved through each technique.

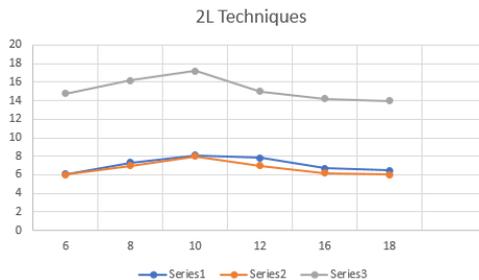

**Figure 3: Result of 2L Technique [9]**

These techniques can eradicate the volume of memory for loading genomic material [9] and another indication is a totally innovative method which is entirely different and founded on observing a DNA base pair arrangement as a symmetrical issue. It is joined with the N-grams procedure too. For increasing the efficiency of information system and dropping the volume of required space, finding recurrent arrangements is a basic step [10]. According to equivalent patterns, Refining List is generated as represented in Figure 4.

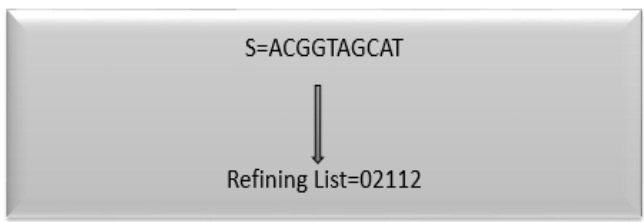

**Figure 4: Refined Subsequence [9]**

If there is a competition at the existing location Authors cross scrutinize the Refined List of their design with the quantity of DNA arrangement Refined List that matches to the presence of the design in the classification. If there is a competition, then this design certainly seems in original DNA arrangement at this existing location.

W. De Neve has proposed a new method for DNA read compression using machine learning performances. The authors described that how to manage storage volume for large amount of DNA sequence data. The goal of their work was to accomplish widespread examination on whether it is practicable to brand a changeover to machine learning methods and additional precise deep learning methods.

Zhao-Hui Qi et al, proposed Huffman coding algorithm to visualize and evaluate DNA classifications in 2011. In this work, authors described about Long chain of DNA representation in the shape of a binary tree. Four nucleotides of DNA as A, C, T, G can be described using the graphical method by assigning them a specific code. Codes are assigned to the bases of DNA according to their frequencies. Figure 5 shows that there will be no loss of information while representation of genes sequence in the shape of binary tree [11].

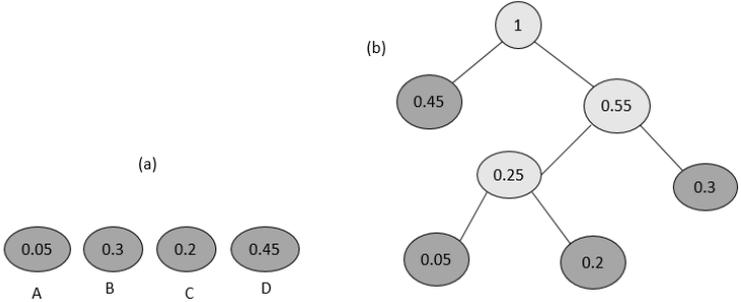

**Figure 5:** A binary tree of a DNA sequence frequency (0.05; 0.3; 0.2; 0.45); a) The primary nodes; b) Huffman tree [11].

Multivariable component is used to measure the quantity of nucleotides that are derived from graph. Graphical approaches have appeared as an influential means for the examination and visualization of extensive DNA classifications. The inspection of resemblances and differences between the comprehensive encoding arrangements of b-globin genetic factor of 11 classes and 6 ND6 proteins demonstrates the efficacy of the system [12].

Vilas Machhi1 et al, proposed compression techniques applied to DNA data of various species in 2016. Authors had worked on four algorithms in order to compress the DNA classification data. (Lampel ziv Welch) LZW algorithm, run length instruction procedure, Mathematics coding and Replacement method. The density outcomes on these processes are offered and compared on DNA categorization data of 10 unalike types [13]. LZW is a dictionary-based compression procedure. It encrypts Deoxyribonucleic Acid data founded on dictionary instead of tabularizing character counts and building tree like Huffman coding Algorithm. For encrypting a substring, a solo encryption number matching to that substring's directory in the dictionary requires to be printed to the resulted file. It provides best output on files with recurrent substrings like text files. Here is a ratio of compression

Table 3: Comparison of Compression Results on h3n2 Virus [13]

| Algorithm | Size | Compression ratio |
|---|---|---|
| Gzip | 101,494 | 94.92% |
| Bzip | 54,314 | 97.28% |
| DNA Compress | 474,139 | 76.26% |
| Gene Compress | 25,837 | 98.705 |
| Binary | 73,901 | 96.305 |
| Huffman | 52,927 | 97.35% |

Markus Hsi-Yang Fritz et al, proposed efficient storage of high throughput DNA sequencing data using reference-based compression in 2017. In this article, authors presented a novel reference-based firmness technique that proficiently compresses DNA bases for loading. According to this technique firstly they line up the new bases arrangements to a references genome after the alignment they encode the dissimilarities among the new bases and existing reference genomes for loading in a memory.

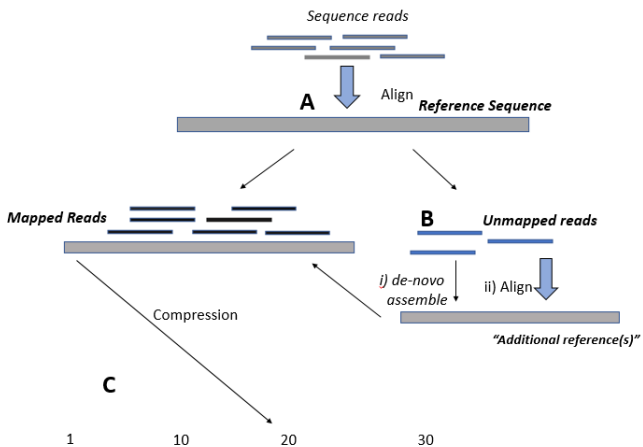

```
                    1        10        20        30
ACGATCTTAATGCCTTACTTGTT- - GG - CATTC        reference
ATCGTAAT---TTAC
        ATGCCTTACTTGT                        mapped reads
              ACTTGTTATGGCC
```

| Position | Strand | Substitutions | Insertions | Deletions |
|---|---|---|---|---|
| 4 | + | 4-G | none | 5-3 |
| 6 | + | none | none | none |
| 7 | + | none | 8-AT 4-C | none |

(A). reads aligned reference, (B). Unaligned are pooled to "compression framework", (C). information of base pair is stored in specific offsets, with insertions, substitutions, or deletions [2].

In the storage of valuable material and unaligned classifications for measurable loss of information their compression algorithm is more efficient.

With the new methods of compression, we can find out the exponential efficiency by increasing the read lengths, and by changing the volume of valuable information magnitude of this competence improvement can be measured. [2].

David Loewenstern et al, introduced new technique of DNA sequence compression which is named as DNA Sequence Classification Using Compression-Based Induction in 2014. Models are formed for using such methods on DNA base pair identification problems that are dependent on the exact positions of genes and assume the individuality of successive gene positions. This work describes as a compression-based induction (CBI), it is geared to the classification learning methods as that raise during learning DNA classification. Because of these algorithms new models are formed that are founded on the main relative positions of genes and on dependence of successive positions [6].

Sheng Bao et al, projected a method of DNA compression in 2005 that is known as A DNA Sequence Compression Algorithm Based on LUT and LZ77. The kernel of compression is a plotting among base folder and terminus folder, the compression method offers to discover the connection. Shared a LUT-based precoding routine and LZ77 compression procedure, this method can achieve a looseness ratio of 1.9bits /base and may be less than this.

Table 4: Experimental Result on File Size [14]

| Sequence | Base number | File size (our algorithm) | File size(GZip) | R(bits/base) |
|---|---|---|---|---|
| Atatsgs | 9647 | 18736 | 20936 | 1.9422 |
| atef1a23 | 6022 | 11448 | 12272 | 1.9010 |
| Atrdnaf | 10014 | 20256 | 22816 | 2.0228 |
| Atrdnai | 5287 | 10192 | 9964 | 1.9277 |
| Chmpxx | 121024 | 237744 | 276192 | 1.9644 |
| Chntxx | 155939 | 309256 | 364104 | 1.9832 |
| Hehcmvcg | 229354 | 466296 | 533888 | 2.0331 |
| HSG6PDGEN | 52173 | 102296 | 117096 | 1.9601 |
| HUMDYSTROP | 38770 | 77504 | 91624 | 1.9991 |
| HUMHDABCD | 66495 | 118424 | 131848 | 1.7809 |
| **Average** | -- | -- | -- | **1.9494** |

The main benefit of this method is fast implementation, less memory space and easy execution [14]. Each participating sign looks up its consistent production which is deposited in the chip before. In this paper, authors built a limited Look-Up Table which executes the plotting association of coding procedure [15]. Wenjing Fan et al, proposed a Complementary Contextual Models with FM-index for DNA Compression in 2017. They presented an effectual reference-based method for DNA arrangement looseness that integrates FM-index and complementary context illustration to improve the effectiveness. GReEn [16] modified the probabilistic sample for mathematical programing founded on the circulation of models. A two-pass looseness basis COMPACT [17] was industrialized to compress matchless cyphers founded on the complementary contextual models. But if the sequences are not well aligned then it may fail. self-index algorithm [18] was proposed by Ferragina and Manzini which has a leveraged relationship among the suffix array data structure [19] and Burrows Wheeler Transform [6] that is used to gain a competent substring penetrating in the space of the warehoused text. To establish a smallest, read classification BWT-Based density method was introduced by Kuma et al [20].

Shakeela Bibi, Javed Iqbal, Adnan Iftekhar, Mir Hassan- **Analysis of Compression Techniques for DNA Sequence Data**FM-index introduced in this paper is only compatible for smaller patterns which are similar, and it is used in [21]. In the first step of this method largest similar values are located by using the converse of reference index in an effective manner. Self-Indexing is use for compression and storing the converse index of reference as a reference arrangement for coding and decoding. Output of Experiments show the introduced algorithm outstrips the state-of-the-art in the density ratio [22].

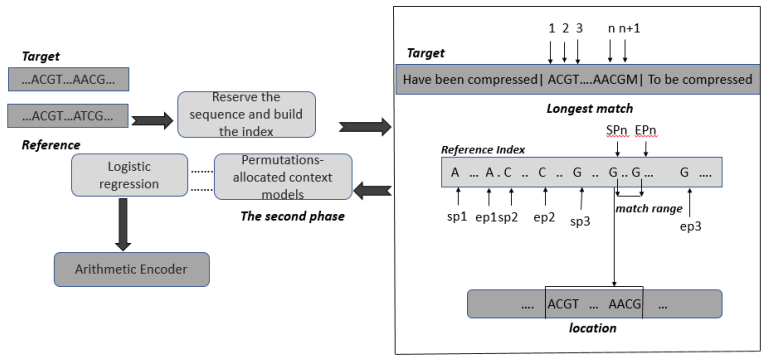

Figure 6: The framework of the introduced algorithm. In the primary step, the variable-length substring is matched and located using the reference index. The remaining unencoded symbols are compressed by the synthesis of complementary contextual models in the second pass [22].

## III.     RESEARCH PROBLEM AND OBJECTIVES

The rapid growth of DNA structures in the DNA collection has been one of the encounters for bio scientists to examine accurately the big capacity of genomic DNA structure data. The firmness of DNA orders is founded on text solidity procedures. However, researchers proved that conservative text solidity methods are inadequate for DNA orders compaction. Therefore, they anticipated special density methods.

In this work, we are addressing the main problems and advantages of DNA data compression since, DNA sequence data is too large to compare, manage and analyze. So, our main objective is to compress the DNA and protein sequence data in such a way that there should not be any loss or degeneracy of Shakeela Bibi, Javed Iqbal, Adnan Iftekhar, Mir Hassan- **Analysis of Compression Techniques for DNA Sequence Data**

DNA sequence information. By reducing size of data whole information about DNA would be easy to store, manage and analyze. Sometimes, DNA sequence may also contain redundant data. The implemented algorithm will also remove such data redundancies. This would be beneficial for bio scientists to manipulate biological data easily and to reduce the dimensions of biological data.

## IV. RESULTS AND DISCUSSION

Since DNA has long chain of four bases (A, G, T, C) with variable length sequence ranging from few bases to hundred. The complex nature of DNA sequence create challenges to manage this huge amount of data. To meet these challenges, several types of techniques have been developed recently and few of them has been explored deeply in this paper. From the comparison of different techniques we have found that different types of techniques show different results for DNA sequences compression. As in [5] Run-Length Encoding Algorithm is used to compress the datasets of HUMHBB (Human Globin), HUMHD ABCD (Human Sequence of contig), MPOMTCG (mitochondrial gene) and VACCG (vaccinia virus genome). Results of this method indicates that the compaction ratio is smaller than 1 BpB and the volume of sequences is condensing to 25% of its original volume. But, there are some limitations of this technique such as this technique is efficient only for limited datasets and not useful for other than Huffman coding algorithm that are based on statistical approaches. While in [9] N-grams/2L Palindrome algorithm is used for compression, which converts N characters into N/2 DNA character set means it shows 50% compression ratio. But this technique is very difficult to implement. At the end, the magnitude of index becomes larger and the efficiency of queries becomes so slow. N-grams/2L algorithm offers space efficiency only on frontend.

Zhao-Hui Qi et al, proposed Huffman coding method to visualize and analyze DNA sequences in 2011. Huffman Coding Algorithms compress the B-globin gene of 11 distinct species Six ND6 protein without any loss of information. In [13] many techniques have been discussed that shows different results as LZW gives 45.1% compression ratio for H3N2, whereas E-coli, Gzip and Bzip gives 73.1%, 94.3% and 97.23% respectively. In [2], Raw FASTQ gives 19.96 efficiency for NA12878 chrom20, Raw FASTA is used for compression that gives 11.45 (for NA12878 chrom20) and Bzip2 FASTQ gives 6.64 (for NA12878 chrom20). These techniques have some drawbacks as it involves a control loss of accuracy in terms of classification information and there is a tradeoff between efficient storage and lose of information.

      It was also found that the illustration through Huffman coding is inimitable when a Huffman compression is well-defined by the incidence distinguishing of base A, T, G, and C in given DNA orders. The key gain of this method is that it permits to have a density ratio per base inferior than 0.8 BpB thus improved than all current density systems. The procedure is also informal to contrivance and stimulating to custom as the five systems bandage the original nucleotide illustration to less than 25%.

**Table 3: Comparison of Sequence Compression Techniques**

| Ref | Techniques | Results | Dataset(s) | Complexity | Drawbacks |
|---|---|---|---|---|---|
| [1] | Run-Length encoding Algorithm | 1. Ratio is smaller than 1 BpB for firmness of genomes mentioned in a paper. 2. volume is condensed to smaller than 25% of its original. | 1. HUMHBB (Human Globin) 2. Human Sequence of contig 3. mitochondrial gene 4. vaccinia virus genome | Low complexity | 1. Efficient for limited datasets. 2. Not useful for other than Huffman coding algorithm that are based on statistical approaches. |
| [2] | N-grams /2L Palindrome Algorithm | 1. size of string having N characters change into N/2. 2. 50% compression ratio. | Unspecified | High complexity (Palindrome Algo) | 1. The magnitude of index becomes larger and the efficiency of queries becomes so slow. 2. N-grams |

Shakeela Bibi, Javed Iqbal, Adnan Iftekhar, Mir Hassan- **Analysis of Compression Techniques for DNA Sequence Data**

| | | | | | technique is difficult to implement. 3.N-grams/2L algorithm offers space efficiency only on frontend. |
|---|---|---|---|---|---|
| [3] | Huffman Coding | Lossless compression of long DNA chain. | B-globin nucleotides of 11 different classes 6-ND6 protein. | O(nlogn) | Only 4 bases of DNA (ACTG) can be map graphically not Protein and RNA. |
| [4] | 1.LZW 2.Gzip 3.Bzip 4.RLE 5.Arithmatic | 1.H3N2 has 45.1% compression ratio. 2.E-coli has 73.1% compression ratio. 3.Gzip gives 4.3% compression ratio for H3N2. 4.Bzip gives 97.23%. | 1.H3N2 2.E-coli 3.Becteria 4.Tomato 5.Rabbit | O(n) | 1.Run length algorithm is not suitable for DNA data compression. |
| [5] | 1.Reference based compression 2.Raw FASTQ 3.Bzip2FASTQ 4.Bzip2 5.BAM 6.Bzip2 clipped | 1. Raw FASTQ gives 19.96 efficiency for NA12878 chrom20. 2. Raw FASTA gives 11.45 (for NA12878 chrom20). 3. Bzip2 FASTQ gives 6.64 (for NA12878 chrom20). | 1.NA12878 CHROME20 2.Pathovar syringas B728a | Unspecified | 1.Require a control loss of accuracy in terms of classification material stowed. 2.Tradeoff between efficient storage and lose of information. |
| [6] | 1.Compression based induction 2.Zdiff, LZ78 3. basic-cbi Tryeach-hamming | Unspecified | 1.ECORGNB 2.ECORPLRPM 3.EORRNH | High complexity | 1. Difficult to understand. 2.More time to compress 3.Due to biased in negative training data, the estimated true error rate was less than 0.1 %. |
| [7] | LUT & LZ77 | 1. Compression is 75%. 2. Ratio of compression is 0.2bits/base larger than theirs commonly. | 1.Sequence: atatsgs 2.HSG6PDGEN 3.HUMDYSTROP 4.HUMHDABCD 5.Mtpacg 6.Mm2p3g | 10^-3 seconds | 1.Time cost by LZ77 isn't so beneficial. 2.LUT pre-coding can't be combined with other distinct compression methods, to review this procedure to recover its efficiency. |

Compression based induction, Zdiff, LZ78 and basic-cbi Tryeach-hamming are used for compression in [6] that shows different results as compared to above mentioned DNA sequence compression methods, In these compression algorithms, ECORGNB, ECORPLRPM and EORRNH datasets are used. But these techniques are difficult to understand, with more time to compress and due to biased in negative training data. The estimated true error rate was less than 0.1 % [6].

At the end in [14] LUT (Look-Up Tables) & LZ77 are discussed. It uses the datasets of Sequence as atatsgs, HSG6PDGEN, HUMDYSTROP, HUMHDABCD Mtpacg and Mm2p3g. It reduces the volume of datasets into 75% of the one that is only reduced by pre-coding routine and ratio for compression of this method is almost 0.2bits/base larger than commonly used methods. There are some drawbacks of these two techniques as the time cost by LZ77 is not so beneficial and LUT pre-coding can't be merged with other distinct compression methods, to review this process to increase its efficiency. Moreover, the analysis of literature also shows that HUFF+ASCII time execution time is less than CTW+LZ sophisticated than DNA Pack and DNA Compress [5]. For further reduction of time execution, it is possible to parallelize its efficiency.

## V.    KEY FINDINGS

Biological data (DNA, RNA, Protein,) compression is very helpful tool to recuperate the useful information from biological sequence. Many techniques has been developed for compression of DNA sequence data, but all have some drawbacks as some techniques are not efficient for large datasets, some are not good for statistical analysis and some of them has more running time. On the other hand, implementation of these techniques is very much complex which takes considerable time to process datasets. Most of the techniques provide best compression ratio but in our opinion Huffman Coding is the best method to compress DNA sequence data which has binary classification tree for compression, codes are assign to the four bases of DNA suppose as A=00, C=01, G=10, T=11, long chain of DNA sequence can be graphically represented which is easy to understand and main benefit of Huffman Coding is that it Shakeela Bibi, Javed Iqbal, Adnan Iftekhar, Mir Hassan- **Analysis of Compression Techniques for DNA Sequence Data** provides lossless compression. Occurrence frequencies of DNA bases are represented in the form of binary trees. For measurement of the quantity of DNA nucleotides datasets multivariable components are used which are derived from graphical representation of long chain of sequences. Huffman Coding is easy to implement and understand but this technique also has a drawback that it is implemented in literature only for DNA nucleotide datasets. Huffman Coding can be applied for protein sequences for efficient compression of datasets. Furthermore, novel graphical representation methods can be proposed in the future for efficient and simple understanding of this complex structured sequence data. While working on any such methods, the semantics of information should not be lost in any case. The existing techniques also fails to consider particular structure of the mentioned sequence data which might be useful for storing sequence information in a precise manner.

## VI. CONCLUSION

The DNA sequence statistics is used in forensics and investigation of drug and for pedigree. The efficient storage of DNA sequences would be extremely helpful in the advance research in biological data mining domain. This work critically analyzes various techniques of DNA data compression and efficient storage. DNA database extent will endure to surge with the amplified sequencing labors all over the world. This hassles the requirement for well-organized data storage approaches. The Huffman coding technique suggests advanced effectiveness particularly for large amount of DNA sequence data but this method is a bit slow and also have some performance overheads. The overall analysis indicates that efficient DNA sequence analysis and compression is still remains an open challenge for the research community. In future, the existing methods of DNA compression may be further enhanced due to their potential applications. Additionally, we may attempt to subordinate these methods to other density processes grounded on statistical methods to bandage the DNA classifications with a rate sophisticated than the proportion of previous methods.

Shakeela Bibi, Javed Iqbal, Adnan Iftekhar, Mir Hassan- **Analysis of Compression Techniques for DNA Sequence Data**